\documentclass[acmsmall,screen]{acmart}

\newcommand{\runnableRepoCount}{249\xspace}
\newcommand{\unrunnableRepoCount}{91\xspace}

\newcommand{\runmutation}{51\xspace}

\newcommand{\numMutators}{16\xspace}

\newcommand{\coverageAllMedian}{86.7}
\newcommand{\coverageAllPercentileTwentyFive}{66.5}

\newcommand{\coverageStMedian}{59.6}
\newcommand{\coverageNostMedian}{69.9}
\newcommand{\coverageUniqueStMedian}{4.4}
\newcommand{\coverageUniqueNostMedian}{10.8}

\newcommand{\mutationAllMedian}{36.5}

\newcommand{\mutationStMedian}{22.2}

\newcommand{\mutationNostMedian}{28.8}

\newcommand{\mutationEffect}{1.3}

\newcommand{\coveredMutationAllMedian}{68.4}
\newcommand{\coveredMutationStMedian}{52.2}
\newcommand{\coveredMutationNostMedian}{52.8}

\newcommand{\benefitEaseOfImpl}{53.3}
\newcommand{\benefitPreventRegressions}{28.3}
\newcommand{\benefitTransparency}{26.7}

\newcommand{\effectiveProgramAnomaly}{70.0}
\newcommand{\effectiveGUIIssues}{53.3}
\newcommand{\effectiveTestCodeIssues}{36.7}
\newcommand{\effectiveConfigureIssues}{26.7}

\newcommand{\challengeReviewingLargeResults}{28.3}
\newcommand{\challengeFalsePositives}{21.7}

\newcommand{\mutationUniqueStMedian}{9.5}             
\newcommand{\mutationUniqueNostMedian}{9.9}           
\newcommand{\mutationCommonMedian}{12.4}               

\newcommand{\surveyRespondents}{60}
\newcommand{\surveyTotalSelections}{134}
\newcommand{\rqFourProgramAnomalyCount}{42}
\newcommand{\rqFourProgramAnomalyPct}{70.0}
\newcommand{\rqFourGUICount}{32}
\newcommand{\rqFourGUIPct}{53.3}
\newcommand{\rqFourTestCodeCount}{22}
\newcommand{\rqFourTestCodePct}{36.7}
\newcommand{\rqFourConfigCount}{16}
\newcommand{\rqFourConfigPct}{26.7}
\newcommand{\rqFourPermissionCount}{3}
\newcommand{\rqFourPermissionPct}{5.0}
\newcommand{\rqFourPerformanceCount}{3}
\newcommand{\rqFourPerformancePct}{5.0}
\newcommand{\rqFourDatabaseCount}{3}
\newcommand{\rqFourDatabasePct}{5.0}
\newcommand{\rqFourSecurityCount}{2}
\newcommand{\rqFourSecurityPct}{3.3}
\newcommand{\rqFourNetworkCount}{1}
\newcommand{\rqFourNetworkPct}{1.7}
\newcommand{\rqFourOtherCount}{10}
\newcommand{\rqFourOtherPct}{16.7}

\newcommand{\implement}{32}

\newcommand{\regression}{17}

\newcommand{\checkresult}{16}

\newcommand{\seriarizable}{10}

\newcommand{\benefitSerializablePct}{16.7}
\newcommand{\update}{10}

\newcommand{\benefitEasyToUpdatePct}{16.7}
\newcommand{\maitainability}{3}

\newcommand{\benefitImproveMaintainabilityPct}{5.0}
\newcommand{\refactoring}{2}

\newcommand{\benefitEasierRefactoringsPct}{3.3}
\newcommand{\AltUItest}{2}

\newcommand{\benefitAlternativeToUITestsPct}{3.3}
\newcommand{\implementPercentage}{53.3\%}
\newcommand{\regressionPercentage}{28.3\%}
\newcommand{\checkresultPercentage}{26.7\%}

\newcommand{\validRespondetDisadvantage}{45}

\newcommand{\hardToReview}{17}

\newcommand{\challengeHardToReviewPct}{28.3}
\newcommand{\falsePositives}{13}

\newcommand{\challengeFalsePositivesPct}{21.7}
\newcommand{\unclearIntention}{10}

\newcommand{\challengeUnclearIntentPct}{16.7}
\newcommand{\easyToIgnore}{10}

\newcommand{\challengeEasyToIgnorePct}{16.7}
\newcommand{\flaky}{6}

\newcommand{\challengeFlakyPct}{10.0}
\newcommand{\crossPlatform}{4}

\newcommand{\challengeCrossPlatformPct}{6.7}
\newcommand{\maintainanceCost}{3}

\newcommand{\challengeMaintenanceCostPct}{5.0}
\newcommand{\largeSnapshots}{2}

\newcommand{\challengeLargeSnapshotsPct}{3.3}
\newcommand{\hardToReviewRate}{28.3}
\newcommand{\falsePositiveRate}{21.7}

\newcommand{\benefitsAgreementRate}{93.97}
\newcommand{\challengesAgreementRate}{92.34}
\newcommand{\benefitsKappa}{0.65}
\newcommand{\challengesKappa}{0.46}
\newcommand{\benefitsKappaInterpretation}{substantial}
\newcommand{\challengesKappaInterpretation}{moderate}

\newcommand{\rqa}{$RQ_1$}
\newcommand{\rqb}{$RQ_2$}
\newcommand{\rqc}{$RQ_3$}

\newcommand{\rqA}{\rqa: To What Extent Does Snapshot Testing Enhance Code Coverage?}
\newcommand{\rqB}{\rqb: To What Extent Does Snapshot Testing Enhance Mutation Score?}
\newcommand{\rqC}{\rqc: What Are Developers' Perceptions of Snapshot Testing?}
\newcommand{\studiedrepo}{340\xspace}
\newcommand{\sent}{1,159\xspace}
\newcommand{\delivered}{1,159\xspace}
\newcommand{\undelivered}{0\xspace}
\newcommand{\responses}{66\xspace}
\newcommand{\validresponses}{60\xspace}
\newcommand{\invalidresponses}{6\xspace}
\newcommand{\bresponse}{56\xspace}
\newcommand{\cresponse}{60\xspace}
\newcommand{\statement}[1]{
\begin{quote}
\faMale~\textit{``#1''}
\end{quote}
}

\usepackage{fancyvrb}
\usepackage{xurl}
\usepackage{booktabs}
\usepackage{url}

\usepackage{amsmath,amssymb,amsfonts}
\usepackage{algorithmic}
\usepackage{graphicx}
\usepackage{textcomp}
\usepackage{xspace}
\usepackage{subcaption}
\usepackage{tabularx}
\usepackage{balance}
\usepackage{fontawesome}
\usepackage{fp}
\usepackage{multirow}
\usepackage{upquote}

\usepackage{xcolor}
\definecolor{darkgreen}{rgb}{0, 0.5, 0} 
\definecolor{whitesmoke}{rgb}{0.99, 0.99, 0.99} 
\definecolor{jsred}{HTML}{CF222E}     
\definecolor{jspurple}{HTML}{8250DF}  
\definecolor{jsstring}{HTML}{0A3069}  
\definecolor{jsblue}{HTML}{0550AE}    
\definecolor{jscomment}{HTML}{6E7781} 
\definecolor{jsframe}{HTML}{333333}   

\definecolor{markercolor}{HTML}{F0E0B0}  

\newcommand{\lsthl}[1]{{\setlength{\fboxsep}{1pt}%
  \colorbox{markercolor}{\strut\ttfamily\footnotesize\color{jspurple}#1}}}

\def\Underline{\setbox0\hbox\bgroup\let\\\endUnderline}
\def\endUnderline{\vphantom{y}\egroup\smash{\underline{\box0}}\\}
\def\|{\verb|}

\newcommand{\ie}{\textit{i.e.,}\xspace}
\newcommand{\eg}{\textit{e.g.,}\xspace}

\newcommand{\etal}{\xspace\textit{et al.}\xspace}

\newcounter{findings_no}

\usepackage{listings}
\definecolor{backcolour}{rgb}{0.95,0.95,0.92}
\lstdefinelanguage{diff}{
  morecomment=**[f][\color{red}]{-},         
  morecomment=**[f][\color{darkgreen}]{+},       
  moredelim=**[is][\bfseries]{@@}{@@},
}
\definecolor{backcolour}{rgb}{0.95,0.95,0.92}
\lstdefinelanguage{commit}{ 
  breakindent = 0pt,
  numbers=none,
  backgroundcolor=\color{white},
  frame=single,
  xleftmargin=3.5em,
  numbersep=0em,
  xrightmargin=1.5em,
}

\renewenvironment{quote}%
{\list{}{%
\rightmargin=10pt%
\leftmargin=20pt%
}\item\relax}{\endlist}

\usepackage[many]{tcolorbox}  
\tcbuselibrary{listings,breakable}
\definecolor{main}{HTML}{D0D3D4}    
\definecolor{sub}{HTML}{D0D3D4}     
\newtcolorbox{dbox}{
    left=2pt,right=2pt,top=2pt,bottom=2pt,
    enhanced, 
    boxrule = 0pt,
    enlarge top by=5pt,
    enlarge bottom by=3pt,
  }
\tcbset{
    sharp corners,
    before skip = 0.1cm,    
    after skip = 0.01cm      
}

\let\oldcite\cite
\renewcommand{\cite}[1]{~\oldcite{#1}}

\let\oldref\ref
\renewcommand{\ref}[1]{~\oldref{#1}}

\def\summarybox#1#2{
\medskip
\begin{tcolorbox}[
  enhanced,
  title=#1,
  colframe=darkgray,
]
    #2
\end{tcolorbox}
\bigskip
}

\AtBeginDocument{%
  \providecommand\BibTeX{{%
    \normalfont B\kern-0.5em{\scshape i\kern-0.25em b}\kern-0.8em\TeX}}}

\setcopyright{acmlicensed}
\copyrightyear{2026}
\acmYear{2026}
\acmDOI{XXXXXXX.XXXXXXX}

\acmISBN{978-1-4503-XXXX-X/18/06}

\begin{document}

\title{Enhancing Test Effectiveness with Snapshot Testing: An Empirical Study} 
\title{On the Effectiveness of Snapshot Testing: An Empirical Study}



\author{Shun Fujita}
\email{fujita.shun.fp9@naist.ac.jp}
\orcid{0009-0009-4234-0330}
\affiliation{%
  \institution{Nara Institute of Science and Technology}
  \city{Ikoma}
  \country{Japan}
}

\author{Satoru Tomizawa}
\email{tomizawa.satoru.tx9@naist.ac.jp}
\orcid{0009-0001-2210-0569}
\affiliation{%
  \institution{Nara Institute of Science and Technology}
  \city{Ikoma}
  \country{Japan}
}

\author{Yutaro Kashiwa}
\affiliation{%
  \institution{Nara Institute of Science and Technology}
  \city{Ikoma}
  \country{Japan}
}
\email{yutaro.kashiwa@is.naist.jp}
\orcid{0000-0002-9633-7577}

\author{Bin Lin}
\affiliation{%
  \institution{Hangzhou Dianzi University}
  \city{Hangzhou}
  \country{China}
}
\email{b.lin@hdu.edu.cn}
\orcid{0000-0001-6307-8460}

\author{Hajimu Iida}
\affiliation{%
  \institution{Nara Institute of Science and Technology}
  \city{Ikoma}
  \country{Japan}
}
\email{iida@itc.naist.jp}
\orcid{0000-0002-2919-6620}

\renewcommand{\shortauthors}{Fujita, et al.}

\begin{abstract}
Software testing is essential for ensuring quality, yet testing complex front-end user interfaces is often time-consuming and difficult. Snapshot testing has emerged as a low-cost regression testing method that detects unexpected changes by comparing program behavior before and after modifications. Despite its growing adoption, empirical evidence regarding its actual effectiveness remains scarce. 
To fill this gap, this study presents a large-scale empirical investigation into the effectiveness of snapshot testing involving \studiedrepo GitHub repositories and a survey of \validresponses experienced practitioners.

Our quantitative results demonstrate that combining snapshot and traditional non-snapshot tests significantly enhances code coverage, achieving a median statement coverage of \coverageAllMedian\%. 
We find that the two approaches are highly complementary, typically exercising distinct regions of the production code.
Furthermore, test suites incorporating snapshot tests detected \mutationEffect~times more mutants than those relying solely on traditional tests. 
Our qualitative insights from the developer survey reveal that practitioners value snapshot testing for its ease of implementation, ability to prevent regressions, and transparency in checking results. 
However, reviewing large result sets and managing false positives remain significant hurdles. These findings guide the strategic integration and future improvement of snapshot testing in modern development pipelines.



\end{abstract}



\maketitle

\section{Introduction}
Software testing is crucial for ensuring software quality. In modern development, as production code expands, developers must create extensive test programs to verify that the software behaves as expected~\cite{DBLP:journals/ese/ZaidmanRDD11}. 
However, this process is time-consuming and requires significant human resources~\cite{SoftwareMeasurementGuidebook1995}. 
The increasing complexity of production code often results in testing that is insufficient due to time constraints or high maintenance costs, even when automated tests are implemented~\cite{SoftwareMeasurementGuidebook1995, DBLP:conf/icsm/VasquezBMP17}.

Numerous studies have proposed various approaches to mitigate these challenges~\cite{DBLP:conf/sigsoft/FraserA11}. 
Fraser and Arcuri~\cite{DBLP:conf/sigsoft/FraserA11} developed EvoSuite\footnote{\url{https://github.com/EvoSuite/evosuite}} that automatically generates test code from Java bytecode. A report~\cite{DBLP:conf/sbst/VoglS0ACP21} shows that EvoSuite achieved 71\% branch coverage and a mutation score of 53\% on average, using a search budget of 120 seconds on 25 classes. 
However, a controlled study~\cite{DBLP:journals/tosem/FraserSMAP15} found no measurable improvement in the number of bugs actually found by developers using such automated test generation tools. 
Fatima\etal\cite{fatimaFlakifyBlackBoxLanguage2023} proposed model-based approaches to enhance the efficiency of test case generation and methods for partial retesting in response to code changes. 

While these test generation approaches are actively researched, their adoption in practice remains limited~\cite{DBLP:journals/tosem/FraserSMAP15}. In contrast, snapshot testing is increasingly used in front-end development by large-scale companies such as Uber and Spotify~\cite{TestingStrategiesMobileNativeFoundationa}. Unlike traditional unit or functional tests, which define the correct behavior of the products, snapshot testing compares the outputs of the current system state against previously saved snapshots. This technique is particularly effective for verifying outputs like UI layouts, API response data, and rendering results. Snapshot testing only detects differences before and after code changes, without verifying whether the current state is correct.

Due to its popularity among practitioners, snapshot testing has been highlighted in recent studies~\cite{DBLP:conf/msr/BuiR23, DBLP:conf/icsm/FujitaK0I23, DBLP:journals/jss/CruzRV23}. Bui and Rocha~\cite{DBLP:conf/msr/BuiR23} collected 686 repositories adopting snapshot testing. They extracted projects using Jest, one of the frameworks that implement snapshot testing, based on GitHub tags, and selected repositories containing files with the snap extension to create the dataset. 
Fujita\etal\cite{DBLP:conf/icsm/FujitaK0I23} investigated the adoption and maintenance of snapshot testing in front-end development, with a particular focus on projects using the Jest framework for JavaScript and TypeScript. They found that projects incorporating both unit and snapshot tests tend to have a higher number of test cases. Additionally, a significant portion of commits (8.2\%) involve updates to snapshot files, indicating frequent co-evolution with production code. 
Cruz \etal\cite{DBLP:journals/jss/CruzRV23} conducted a grey literature review, revealing both the advantages (such as ease of implementation) and drawbacks (such as fragility) of snapshot testing. 
While these studies highlight the use of snapshot testing, the effectiveness of snapshot testing compared to other testing methods remains an open question. 

In this study, we aim to evaluate snapshot testing using standard testing metrics, including coverage and mutation scores. We collect 13,256 GitHub repositories with more than 500 stars and identify \studiedrepo repositories that employ the most popular framework for JavaScript and practice snapshot testing. We compare the coverages and mutation scores when using snapshot tests and/or other tests. We find that the combination of snapshot and traditional non-snapshot tests achieves a median statement coverage of \coverageAllMedian\%, which is significantly higher than using either method alone. Furthermore, test suites incorporating snapshot tests detected over \mutationEffect~times more mutants compared to those without them, reaching a median mutation score of \mutationAllMedian\%. Our results indicate that snapshot and non-snapshot tests are complementary, as they often exercise distinct areas of production code and kill different types of mutants.

In addition, we conducted an online survey to gain insights into the issues that snapshot testing can identify, as well as developers' perceptions of its benefits and challenges. 
Out of \sent survey invitations, we received \responses responses (5.7\% response rate). We found that developers primarily value snapshot testing for its ease of implementation (\benefitEaseOfImpl\%), ability to prevent regressions (\benefitPreventRegressions\%), and transparency in checking results (\benefitTransparency\%). Practitioners reported that snapshot testing is particularly effective at identifying Program Anomaly issues (\effectiveProgramAnomaly\%) and GUI-related issues (\effectiveGUIIssues\%), and can even detect test code-related issues (\effectiveTestCodeIssues\%) that basic unit test assertions might miss. However, practitioners also face significant challenges, most notably the difficulty of reviewing or interpreting large result sets (\challengeReviewingLargeResults\%) and the introduction of false positives (\challengeFalsePositives\%).


\smallskip
{\sf\bf Replication package.} To facilitate replication and further studies, the data and programs used in this study are publicly available on GitHub repository.\footnote{Replication package: \url{https://github.com/sdlab-naist/fujita_snapshot}} 

\lstdefinelanguage{JavaScript}{
  morekeywords={const,let,var,function,return,if,else,for,while,new,
                async,await,import,export,from,as,true,false,null,undefined},
  morecomment=[l]{//},
  morecomment=[s]{/*}{*/},
  morestring=[b]',
  morestring=[b]",
  morestring=[b]`,
  sensitive=true
}

\lstdefinestyle{jsstyle}{
  language=JavaScript,
  basicstyle=\ttfamily\small,
  keywordstyle=\color{jsred},
  stringstyle=\color{jsstring},
  commentstyle=\color{jscomment}\itshape,
  emph={it,describe,expect,renderMinimap,querySelector,mouseDown,mouseMove,
        toMatchSnapshot,toMatchInlineSnapshot,toHaveStyle},
  emphstyle=\color{jspurple},
  literate={=>}{{\textcolor{jsblue}{=>}}}2,
  frame=single,
  escapeinside={(*@}{@*)},
  rulecolor=\color{jsframe},
  framesep=6pt,             
  basicstyle=\ttfamily\footnotesize, 
  numbers=none,                      
  xleftmargin=20pt,                  
  xrightmargin=20pt, 
  framexleftmargin=4pt,     
  framexrightmargin=4pt,    
  aboveskip=\medskipamount,
  belowskip=\smallskipamount,
  upquote=true,
  backgroundcolor=\color{white}, 
  columns=fullflexible,
  keepspaces=true,
  showstringspaces=false,
  breaklines=true,          
  tabsize=2
}




\begin{figure}[t]
    \centering
    \includegraphics[width=0.95\linewidth]{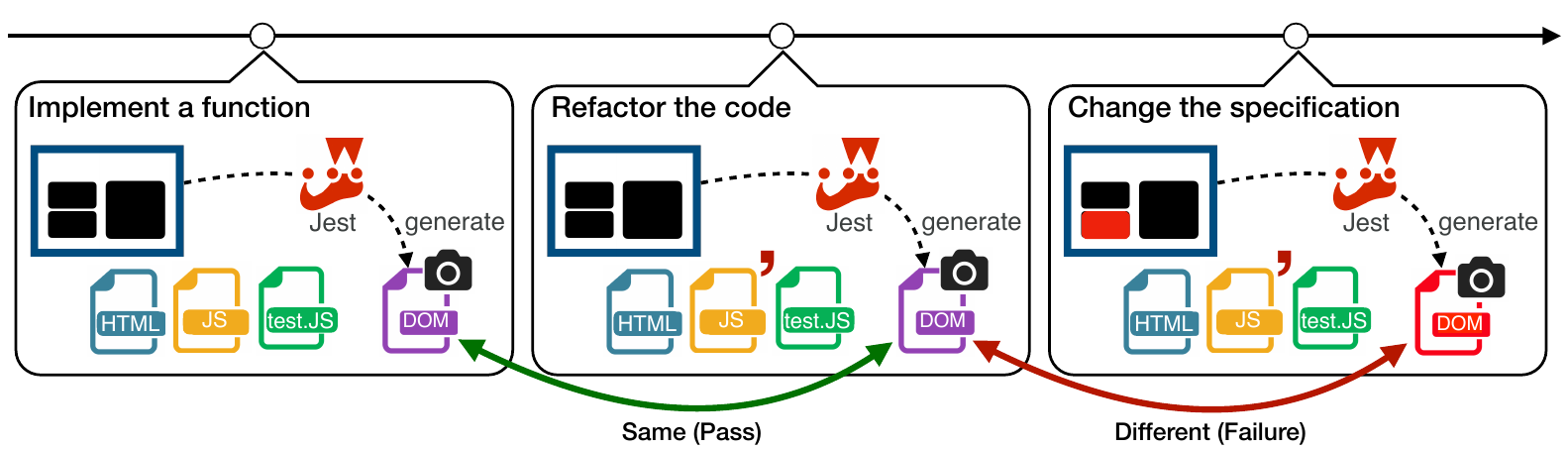}
    \caption{Example use of snapshot tests}
    \label{fig:overview}
\end{figure}
\section{Background: Snapshot Testing with Jest}\label{sec:background}

\begin{figure}[ht]
    \centering
    \includegraphics[width=0.6\linewidth]{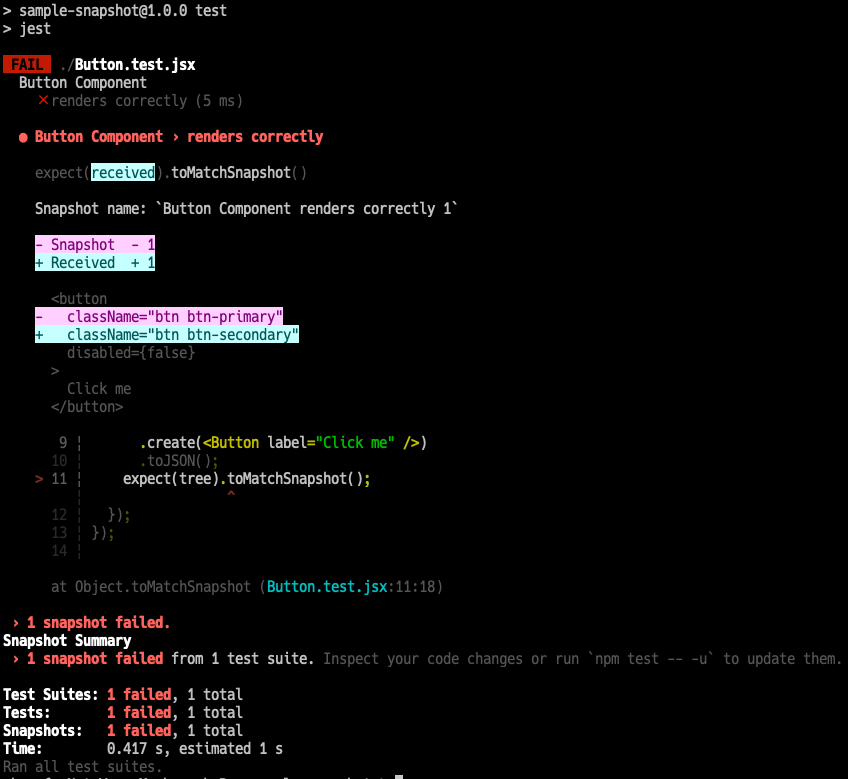}
    \caption{Example of the outputs returned by Snapshot tests}
    \label{fig:output_example}
\end{figure}

Snapshot testing is a type of regression testing 
primarily used to ensure that the behavior of a program, particularly its user interface (UI), remains consistent before and after code modifications.
Figure~\ref{fig:overview} illustrates the overview process of using snapshot testing. 
When snapshot testing is executed for the first time, it saves the current snapshot of the program under test, which could be a serialized version or a screenshot. Subsequently, each time the snapshot test is executed, it compares the current snapshot with the previous one. If there are no changes, the snapshot test passes; otherwise, the test fails and displays the differences. Figure~\ref{fig:output_example} shows an example of outputs that detect a difference from the prior change.

When a snapshot test fails, the developer has two options. The first option is to update the snapshot, which is done when the changes are intentional and as expected by the developer. The second option is to fix the production code to ensure the program under test matches the previous state, which is done when the changes are unintentional.


One of the most popular test frameworks that implements snapshot testing is Jest.\footnote{\url{https://jestjs.io}}
Jest, developed for JavaScript and TypeScript, has over 45,000 stars on GitHub and, as of 2025, is downloaded over 30 million times per week
~\cite{npmtrends2025jest}.
Jest provides two snapshot testing methods, each with its own format to record values such as serialized DOM elements, API outputs, or lists. When a test fails, developers can easily regenerate the snapshot after applying a fix, without needing to modify the test code, thanks to Jest's interactive mode. The two methods Jest offers for snapshot testing are:

\begin{itemize}

\setlength{\itemindent}{-10pt}
    \item {\texttt{toMatchSnapshot()}: } Figure~\ref{fig:file_snapshot_tests} shows an example snippet using this method, and Figure~\ref{fig:file_snapshot_file} depicts an example of snapshot files generated. \texttt{toMatchSnapshot()} stores a snapshot under the ``\texttt{\_\_SNAPSHOT\_\_}'' directory. It then verifies whether the generated snapshots are the same as the ones created previously. If the method is invoked for the first time or the snapshots are deleted, snapshots are only stored, and no verification will be conducted. 
    
    \begin{figure}[!ht]
    \begin{lstlisting}[style=jsstyle][frame=single,xleftmargin=17pt,framexleftmargin=17pt,numbersep=8pt,aboveskip=\medskipamount,belowskip=\smallskipamount]
    it('should position slider controller when performed drag operation on it', () => {
      const { container } = renderMinimap();
      const sliderController = container.querySelector('#sliderController');
      fireEvent.mouseDown(sliderController);
      fireEvent.mouseMove(sliderController);
      expect(sliderController).toHaveStyle('left: 0px');
      expect(container).(*@\lsthl{toMatchSnapshot}@*)();
    });
    \end{lstlisting}
    \caption{Example of snapshot tests using {\small \texttt{toMatchSnapshot()}}}
    \label{fig:file_snapshot_tests}
\end{figure} 
    
    \smallskip
    \item \texttt{toMatchInlineSnapshot(): }
    Figure~\ref{fig:inline_snapshot_tests} shows an example snippet using this method. This method does not store snapshots in files but instead passes the serializable values of the snapshots as an argument, \eg \textit{toMatchInlineSnapshot(``\textless div\textgreater ...\textless /div\textgreater")}. The argument in the method will be automatically modified by Jest when developers update snapshots following Jest's prompt. 
    
\end{itemize}

\begin{figure}[!ht]
\begin{lstlisting}[style=jsstyle]
it('should position slider controller when performed drag operation on it', () => {
  const { container } = renderMinimap();
  const sliderController = container.querySelector('#sliderController');
  fireEvent.mouseDown(sliderController);
  fireEvent.mouseMove(sliderController);
  // The argument below is inserted automatically by Jest
  expect(sliderController.style.left).(*@\lsthl{toMatchInlineSnapshot}@*)(`"0px"`);
});
\end{lstlisting}
\caption{Example of snapshot tests using {\small \texttt{toMatchInlineSnapshot()}}}
\label{fig:inline_snapshot_tests}
\end{figure}

\begin{figure}[!t]
    \centering
        \setlength{\fboxsep}{0pt}%
    \setlength{\fboxrule}{0.5pt}%
    \fbox{ 
    \includegraphics[width=0.55\linewidth]{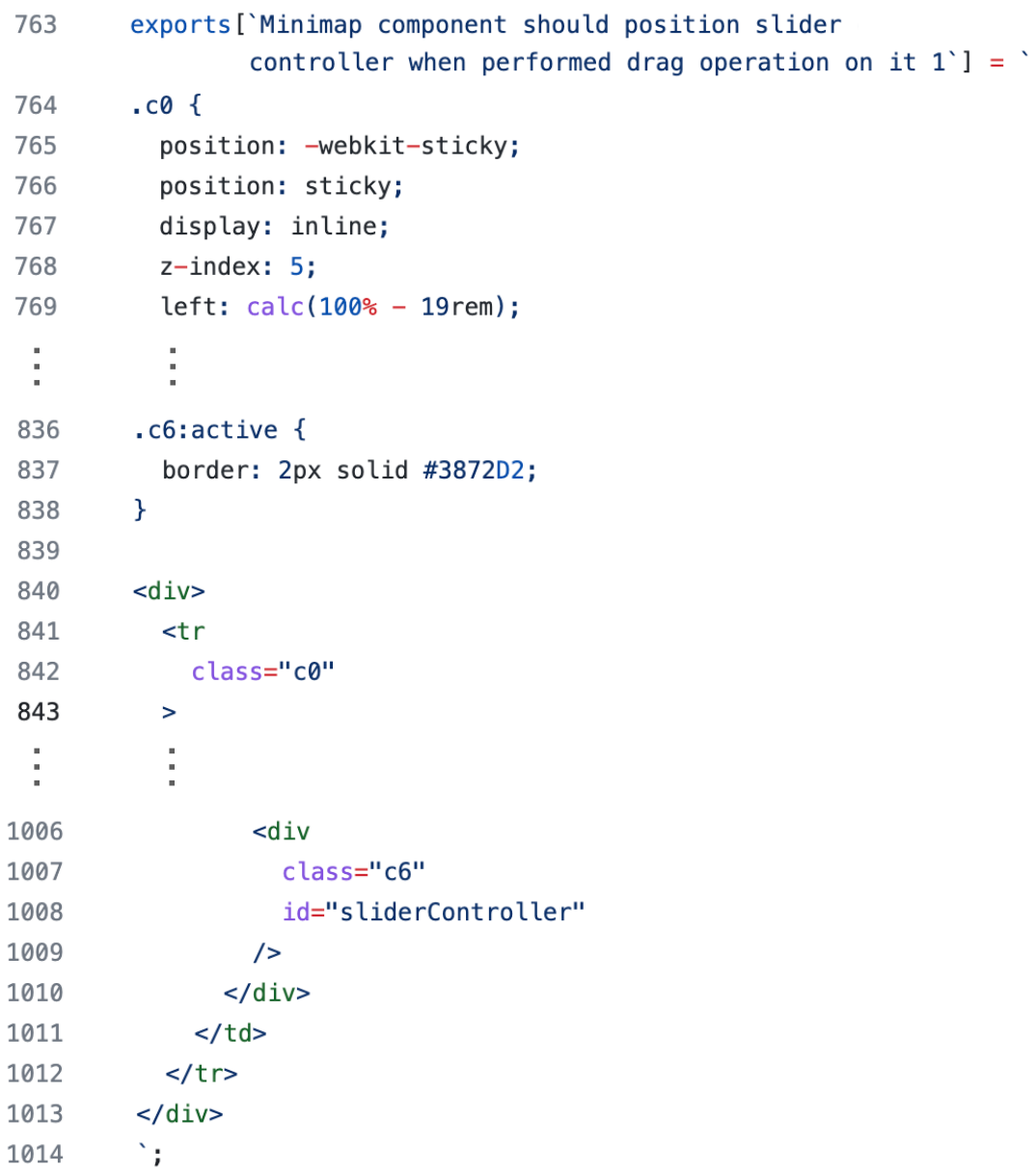}
    }
    \caption{Example of snapshot file made by the test in Figure~\ref{fig:file_snapshot_tests}}
    {\footnotesize This snapshot starts from line 763 and ends at line 1014.}
    \label{fig:file_snapshot_file}
\end{figure}

\newpage
\section{Research Questions}\label{sec:researchquestion}

Our study aims to address the following research questions (RQs):

\subsection*{\rqA}
Code coverage is a widely adopted metric for evaluating the thoroughness of a test suite. Developers often rely on code coverage to gauge the effectiveness and completeness of their tests. High code coverage is commonly seen as an indicator of well-developed and comprehensive testing practices. Many studies~\cite{DBLP:conf/wcre/KochharTL15,DBLP:conf/esem/MockusND09} investigated the effectiveness of coverage and concluded that coverage is correlated with the effectiveness of a test suite. However, it remains unclear to what extent snapshot testing contributes to improving coverage scores. In this research question, we aim to investigate the impact of snapshot testing on code coverage by comparing the coverage achieved with and without snapshot tests.

\subsection*{\rqB}
Mutation testing is another widely used technique for assessing the effectiveness of test suites. It evaluates the quality of software tests by introducing small changes (mutations) to the code and checking if the tests can detect these changes. While code coverage measures the extent to which code is exercised during testing, mutation testing evaluates a test suite's ability to identify potential bugs. Numerous studies~\cite{DBLP:conf/sigsoft/JustJIEHF14, DBLP:conf/icse/PetrovicI18,DBLP:conf/icse/PetrovicIFJ21} have confirmed that the mutation score is a reliable indicator of a test suite's ability to detect bugs, despite the high costs associated with running mutation tests. In this RQ, we compare mutation scores between test suites with and without snapshot tests to determine how effectively snapshot tests identify mutants.

\subsection*{\rqC}
Cruz et al.~\cite{DBLP:journals/jss/CruzRV23} conducted a grey literature review to summarize the advantages and disadvantages of snapshot testing. 
Their review highlighted that snapshot tests are commonly valued for preventing regressions and simplifying test creation, but also noted challenges such as fragility and the potential for bloated snapshot files.
However, the literature on this topic remains limited because snapshot testing has been rapidly adopted by developers. In fact, Cruz et al.'s review includes only 50 reliable sources. In this study, we aim to gain a deeper understanding of developers' perceptions of snapshot testing. 
Specifically, RQ3 is further divided into the following three sub-RQs:
\begin{itemize}
    \item RQ3.1: What types of issues do developers believe can be identified by snapshot testing?
    \item RQ3.2: What advantages do developers perceive in using snapshot testing?
    \item RQ3.3: What challenges do developers associate with snapshot testing?
\end{itemize}


\section{Quantitative Analysis}\label{sec:study_design}
This section presents the design and findings of our quantitative study addressing RQ1 and RQ2.

\subsection{Context Selection}\label{sec:deta_collection}
\begin{table}[t]
\small
\centering
\caption{Repository filtering criteria and resulting dataset sizes}
\begin{tabular}{rlrr}\toprule
                    No.&Filtering& \# repo&\%   \\ \midrule
1&Primarily uses JavaScript&     13,256 &--\\
2&Uses Jest& 1,355 & 10.2\%\\
3&Employs Snapshot testing & \studiedrepo & 2.56\%\\
4&Tests are executable & \runnableRepoCount & 2.26\%\\

\bottomrule
\end{tabular}
\end{table}

We identified repositories using snapshot testing through a systematic filtering process. Jest, one of the most popular testing frameworks in web development,\footnote{\url{https://2024.stateofjs.com/en-US/libraries/testing/}} provides comprehensive snapshot testing support, making it an ideal target for our study.
Using GitHub Search \cite{DBLP:conf/msr/DabicAB21}\footnote{\url{https://seart-ghs.si.usi.ch/}}, we extracted 13,256 repositories that primarily use JavaScript and have more than 500 stars. We then applied the following filtering steps:

\textbf{Step 1: Jest Detection.} We determined whether Jest was being used by checking if the \texttt{package.json} file in the root directory of each repository contains Jest commands in the \texttt{script} property. 

\textbf{Step 2: Snapshot Test Identification.} We executed the command ``\texttt{npx jest --listTests}'' to obtain a list of test files containing test code. We then identified whether snapshot testing was adopted by checking if one or more of the snapshot testing methods were present in the test files. Note that we classify a test as a snapshot test if it uses either \texttt{toMatchSnapshot} or \texttt{toMatchInlineSnapshot}, as these are the only two methods used for snapshot testing in Jest. Tests using other assertions are classified as unit tests. Also, we do not distinguish between unit tests and integration tests. As a result of the data collection process, we identified \studiedrepo repositories that utilize snapshot testing.

\textbf{Step 3: Executability Verification.} We cloned these repositories, executed their test suites using a simple command (\ie npx jest --coverage), and confirmed that \runnableRepoCount repositories can run tests in our environment. With these \runnableRepoCount repositories, we performed the data analysis described in the following section.

\subsection{Data Analysis}
To quantify the contribution of snapshot tests to code quality metrics, we measured coverage and mutation scores under three conditions:
\begin{enumerate}
    \item Running only non-snapshot tests
    \item Running only snapshot tests
    \item Running all tests
\end{enumerate}

The difference between metrics from all tests and non-snapshot tests only represents the contribution of snapshot tests. The metrics from snapshot tests alone indicate how extensively these tests exercise the production code.

To classify whether a test is a snapshot test, we used abstract syntax trees (ASTs). Using \texttt{babel-parser},\footnote{\url{https://babeljs.io/docs/babel-parser}} we parsed test files into ASTs and identified snapshot tests as those containing snapshot methods (\texttt{toMatchSnapshot} or \texttt{toMatchInlineSnapshot}) as descendants of test case nodes (\texttt{test} or \texttt{it}). In this way, we obtain two test subsets: snapshot tests and non-snapshot tests. 

We ran these test subsets using Jest's skip functionality. 
Specifically, we appended the \texttt{.skip} annotation to the test methods (\eg replace \texttt{test} with \texttt{test.skip} and \texttt{it} with \texttt{it.skip}) so that the execution of tests is skipped. 
This modification was performed by parsing test code into ASTs using \texttt{babel-parser}, modifying the relevant nodes, and converting back to source code using \texttt{babel-generator}.\footnote{\url{https://babeljs.io/docs/babel-generator}}
Note that repositories encountering errors during test execution were excluded from our analysis. The number of excluded repositories differed between coverage and mutation testing due to different error conditions.


\subsubsection{\rqa (Coverage)}
Coverage refers to the percentage of the production code that is exercised by the test code, which is often used as a proxy metric to evaluate the quality of tests. A higher coverage percentage indicates that a large portion of the production code is being tested. 
We executed the tests using the Jest command ``\texttt{npx jest --coverage}.'' and collected the coverage data. While Jest supports multiple coverage types (statement, branch, and function coverage), we focused on statement coverage for consistency with common practice. 

\subsubsection{\rqb (Mutants)}

Mutation testing is a technique used to evaluate test code by introducing artificially created bugs, known as mutants, into the production code to determine whether the test code can detect these bugs. This method has been employed to assess tests in the development environment~\cite{DBLP:conf/icse/PetrovicI18} and has been shown to correlate with the actual bug detection capability of tests~\cite{DBLP:conf/sigsoft/JustJIEHF14}. We calculate two types of mutation scores based on four statuses of embedded mutants, as illustrated in Table \ref{table:mutant_states}. The two mutation scores are formulated as follows:

\begin{itemize}
    \item Mutation Score: This metric evaluates the extent to which test cases detect faults that are artificially injected throughout the production code. A higher mutation score indicates that the test suite is effective in catching errors. This score is calculated as the percentage of mutants that are detected (killed) by the test cases. It is given by the formula:
    \[
        \textit{Mutation Score} \\[5pt]
        \hfill =\dfrac{Killed + Timeout}{Killed + Timeout + Survived + No Coverage}
    \]
    \item Mutation Score Based on Covered Code: This score focuses on the mutants that are within the parts of the code actually executed by the test cases. This metric is useful because it excludes mutants in parts of the code that are not covered by any test, providing a more accurate measure of the test suite's effectiveness. The formula is similar but only considers covered code:
    \[
        \textit{Mutation Score Based on Covered Code} \\[5pt]
        \hfill =\dfrac{Killed + Timeout}{Killed + Timeout + Survived}
    \]
\end{itemize}


\begin{table}[t]
\small
\caption{Examples of Mutants by Stryker}
\begin{tabular}{lp{8cm}}\toprule
Mutator & Example \\ \midrule
\texttt{ArithmeticOperator}    & Change \texttt{+} to \texttt{-}, \texttt{-} to \texttt{+} \\
\texttt{ArrayDeclaration}      & Empty the array \\
\texttt{ArrowFunction}         & Replace the body of an arrow function with \texttt{undefined} \\
\texttt{AssignmentOperator}    & Change \texttt{+=} to \texttt{-=} \\
\texttt{BlockStatement}        & Empty the block \\
\texttt{BooleanLiteral}        & Change \texttt{true} to \texttt{false}, \texttt{false} to \texttt{true} \\
\texttt{ConditionalExpression} & Replace condition with \texttt{true} or \texttt{false} \\
\texttt{EqualityOperator}      & Change \texttt{<} to \texttt{<=} or \texttt{>=}, \texttt{===} to \texttt{!==} \\
\texttt{LogicalOperator}       & Change \texttt{\&\&} to \texttt{||} \\
\texttt{MethodExpression}      & Replace \texttt{min()} with \texttt{max()} \\
\texttt{ObjectLiteral}         & Empty the object \\
\texttt{OptionalChaining}      & Remove optional chaining \\
\texttt{Regex}                 & Change \texttt{abc\$} to \texttt{abc} \\
\texttt{StringLiteral}         & Empty the string \\
\texttt{UnaryOperator}         & Change \texttt{-} used as a sign to \texttt{+} \\
\texttt{UpdateOperator}        & Change increment to decrement \\
\bottomrule
\end{tabular}
\label{tab:mutant}
\end{table}
\begin{table}[t]
\small
\caption{Mutant States}
\label{table:mutant_states}

\begin{tabular}{l p{10cm}}
\toprule
\textbf{State} & \textbf{Description} \\\midrule
Killed & At least one test failed when the mutant was embedded \\
Survived & All tests passed despite the mutant being embedded \\
No coverage & The mutant was not detected because it was not covered by any tests \\
Timeout & The tests did not finish due to an infinite loop or similar issue when the mutant was embedded \\
\bottomrule
\end{tabular}
\end{table}

In this study, to measure the mutation score for JavaScript code, we used the tool Stryker\footnote{\url{https://stryker-mutator.io}}. Stryker is one of the most popular mutation testing tools for JavaScript in practice and has been used in previous studies\cite{DBLP:journals/ese/SanchezDMS22,DBLP:journals/tse/TipBS25}. 
StrykerJS provides \numMutators mutator types,\footnote{\url{https://stryker-mutator.io/docs/mutation-testing-elements/supported-mutators/}} summarized in Table~\ref{tab:mutant}; all of them occurred in the repositories we analyzed. Throughout this paper, we refer to them by Stryker's mutator names (\eg \texttt{EqualityOperator}).We initialized Stryker using ``\texttt{stryker init}'' to configure the package manager and output directory, then executed ``\texttt{npx stryker run}'' to measure mutation scores.





\subsection{Results}\label{sec:results}
\begin{figure*}[!t]
    \centering
    \includegraphics[width=.9\linewidth]{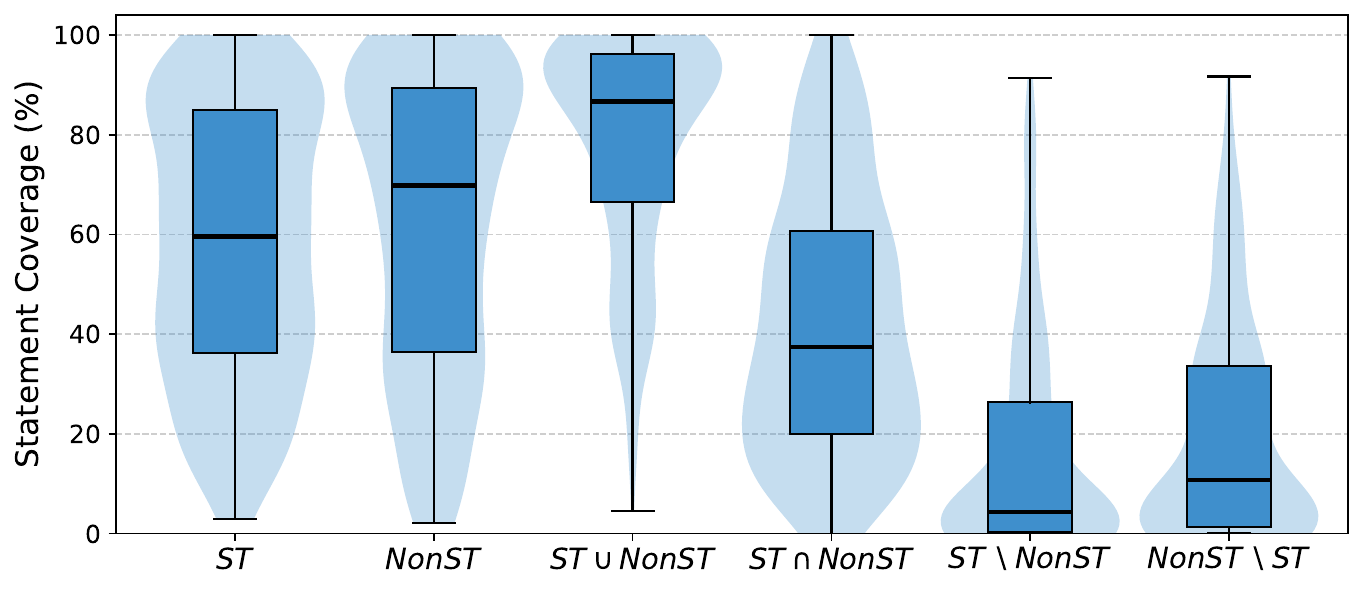}
    
    \caption{RQ1: Distributions of coverages from \runnableRepoCount repositories when running all the tests, only snapshot testing, and only non-snapshot testing.}
    \label{fig:cov_result}

\end{figure*}

\subsubsection{\rqA}\label{sec:result_1}
Out of the \studiedrepo studied repositories, we were unable to run tests on \unrunnableRepoCount repositories in our environment due to the complexity of their testing processes. Consequently, we obtained statement coverage from \runnableRepoCount repositories. Figure~\ref{fig:cov_result} illustrates the distribution of coverages from these \runnableRepoCount repositories when running all tests, only snapshot testing, and only non-snapshot testing. 
In the violin plots, the black line represents the median, the thickness of the outer layer represents the probability density of the plotted values, and the box represents the interquartile range.
$ST$ and $NonST$ denote the coverage obtained by running only snapshot tests and only non-snapshot tests, respectively, and $ST \cup NonST$ denotes the coverage obtained by running all tests. $ST \cap NonST$ denotes the coverage shared by both, and $ST \setminus NonST$ and $NonST \setminus ST$ denote the coverage achieved only by snapshot tests and only by non-snapshot tests, respectively.


The total statement coverage (\ie using both snapshot tests and non-snapshot tests) achieved a median of \coverageAllMedian\%. Additionally, the 25th percentile of statement coverage is \coverageAllPercentileTwentyFive\%, indicating that 75\% of repositories achieve at least this level of coverage.
When comparing the coverage of only snapshot tests and only non-snapshot tests, the coverage of non-snapshot tests (median: \coverageNostMedian\%) is slightly higher than that of snapshot tests (median: \coverageStMedian\%).

However, the total coverage remains significantly higher than that of non-snapshot tests alone (Wilcoxon signed-rank test, $p<0.001$). We then examined the coverage achieved by only one kind of test. The median of $ST \setminus NonST$ is \coverageUniqueStMedian\% and that of $NonST \setminus ST$ is \coverageUniqueNostMedian\%; about two-thirds of the repositories have some coverage unique to each kind of test. This indicates that snapshot tests and non-snapshot tests cover largely the same parts of the production code, while each also covers some areas that the other does not.

\summarybox{\textbf{Answer to RQ1}}{
\textit{The total coverage achieved by combining snapshot and non-snapshot tests is significantly higher than using either method alone, with a median coverage of \coverageAllMedian\%. Non-snapshot tests achieved slightly higher median coverage (\coverageNostMedian\%) compared to snapshot tests (\coverageStMedian\%), but both methods complement each other by covering distinct areas of the production code. This complementary nature highlights the importance of employing both testing methods for comprehensive test coverage.}
}


\subsubsection{\rqB}\label{sec:result_2}
\begin{figure*}[!t]
    \begin{minipage}[b]{0.95\linewidth}
        \centering
        \includegraphics[width=\linewidth]{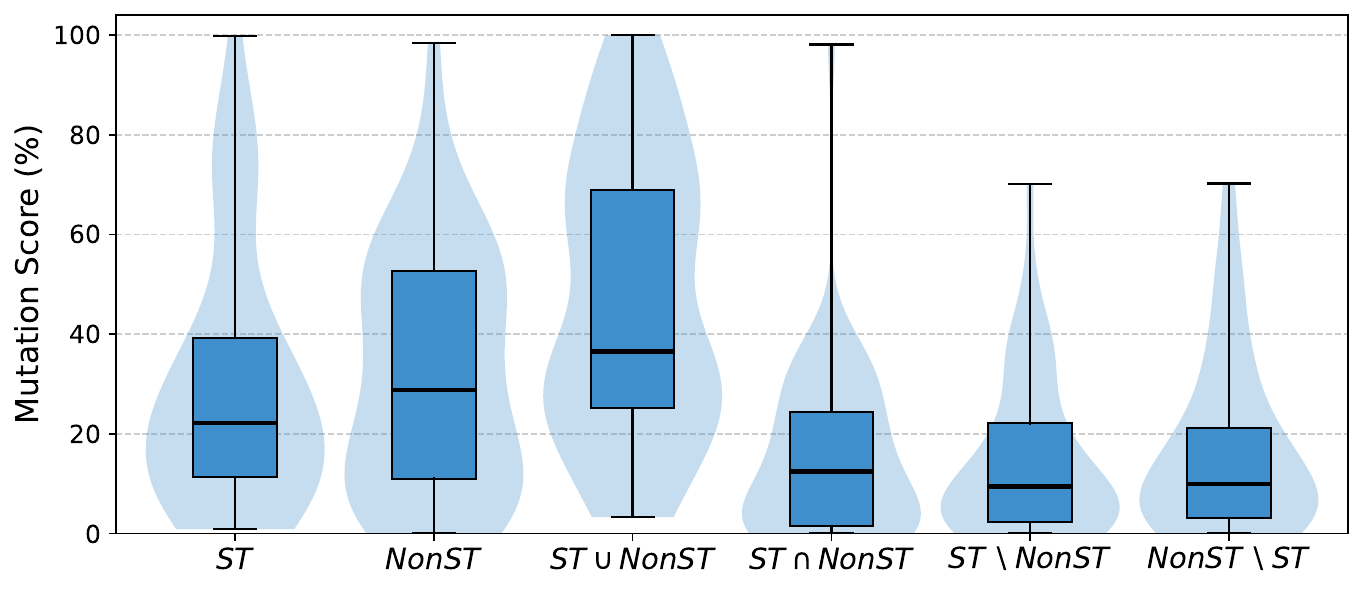}
        \subcaption{Mutation score}
        \label{fig:mutation_score}
    \end{minipage}
    
     \begin{minipage}[b]{0.95\linewidth}
        \centering
        \includegraphics[width=0.5\linewidth]{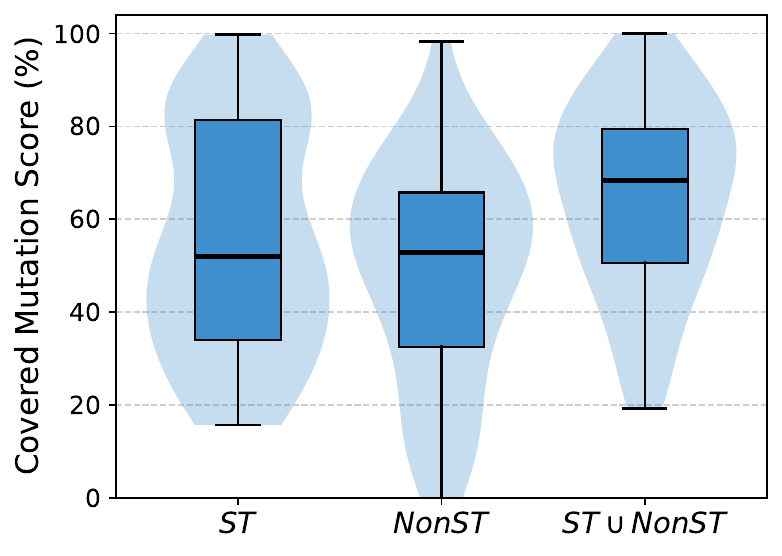}
        \subcaption{Mutation score based on covered code\footnote{Note: We did not compare mutants killed commonly or uniquely across test types, as each type covers different code portions, yielding different sets of target mutants.}}
        \label{fig:mutation_score_coverd}
     \end{minipage}
     \label{fig:mutation}
    \caption{RQ2: Distributions of mutation score obtained from the studied repositories}
\end{figure*}
Of the \runnableRepoCount repositories capable of coverage measurement, we successfully obtained mutation scores for \runmutation repositories. Unlike coverage testing, mutation testing imposes stricter environmental requirements. Specifically, all test cases must pass before mutation testing can proceed, and the source code must conform to the syntax that Stryker supports for mutant insertion.

Figure \ref{fig:mutation_score} illustrates the distribution of mutation scores. For the bare mutation score (\ie without considering coverage), the median across all tests is \mutationAllMedian\%. The median scores for snapshot tests and non-snapshot tests are \mutationStMedian\% and \mutationNostMedian\%, respectively. A Wilcoxon signed-rank test ($\alpha=0.05$) revealed no statistically significant difference between these two groups ($p$=0.750).
Although snapshot tests achieve a slightly lower median mutation score than non-snapshot tests, the combined score is significantly higher than either individual score (Wilcoxon signed-rank test, $p<0.001$).
This is because the two kinds of tests largely kill different mutants: a median of \mutationUniqueStMedian\% of the mutants are killed only by snapshot tests ($ST \setminus NonST$) and \mutationUniqueNostMedian\% only by non-snapshot tests ($NonST \setminus ST$), whereas only \mutationCommonMedian\% are killed by both ($ST \cap NonST$).

We obtained similar results when considering only the covered code. Figure \ref{fig:mutation_score_coverd} shows the distribution of mutation scores calculated over covered lines. The median score for snapshot tests (\coveredMutationStMedian\%) is slightly lower than that for non-snapshot tests (\coveredMutationNostMedian\%), though the difference is not statistically significant. The median score for all tests combined (\coveredMutationAllMedian\%) exceeds both individual scores, with a narrower interquartile range. These results further suggest that snapshot and non-snapshot tests detect different mutants.

\begin{table}[!t]
\centering
\caption{Median kill rates by mutator type for snapshot tests (ST) and non-snapshot tests (NonST), sorted by $\Delta$. $n$ is the number of repositories in which the mutator occurs.}
\label{tab:mutant_type_comparison}
\small
\begin{tabular}{lrrrr}
\toprule
 & & \multicolumn{3}{c}{Median kill rate (\%)} \\
\cmidrule(lr){3-5}
Mutator & $n$ & ST & NonST & $\Delta$ \\
\midrule
\texttt{ArrayDeclaration} & 49 & 20.0 & 14.3 & +5.7 \\
\texttt{ObjectLiteral} & 49 & 24.0 & 20.5 & +3.5 \\
\texttt{StringLiteral} & 51 & 20.2 & 19.4 & +0.8 \\
\texttt{AssignmentOperator} & 17 & 0.0 & 0.0 & +0.0 \\
\texttt{ArrowFunction} & 47 & 29.9 & 30.8 & -0.9 \\
\texttt{Regex} & 32 & 1.7 & 5.8 & -4.1 \\
\texttt{ConditionalExpression} & 51 & 22.4 & 28.6 & -6.2 \\
\texttt{OptionalChaining} & 20 & 3.6 & 11.9 & -8.3 \\
\texttt{EqualityOperator} & 50 & 25.0 & 36.6 & -11.6 \\
\texttt{BooleanLiteral} & 50 & 21.4 & 33.3 & -11.9 \\
\texttt{LogicalOperator} & 50 & 24.7 & 37.1 & -12.4 \\
\texttt{BlockStatement} & 51 & 21.9 & 34.7 & -12.8 \\
\texttt{MethodExpression} & 41 & 3.2 & 16.7 & -13.5 \\
\texttt{UnaryOperator} & 28 & 10.1 & 24.3 & -14.2 \\
\texttt{ArithmeticOperator} & 41 & 1.5 & 31.6 & -30.1 \\
\texttt{UpdateOperator} & 20 & 17.0 & 50.0 & -33.0 \\
\bottomrule
\end{tabular}
\end{table}

When we examine the killed mutants by type, Table~\ref{tab:mutant_type_comparison} compares the kill rates of snapshot and non-snapshot tests for each mutator type. Many repositories have only a few mutants of a given type, so their per-repository kill rates tend to be extreme. For five mutator types, the median kill rate of snapshot tests is below 5\% (0\% for \texttt{AssignmentOperator}). We therefore compare the paired kill rates with two-sided Wilcoxon signed-rank tests. Because we run one test for each of the \numMutators mutator types, the chance of at least one false positive grows with the number of tests, so we adjust the $p$-values using the Holm method to keep the family-wise error rate at 5\%.

After this correction, no mutator type shows a statistically significant difference between snapshot and non-snapshot tests (all $p_{\mathrm{Holm}} > 0.25$). For literal and data-structure mutators such as \texttt{StringLiteral}, \texttt{ObjectLiteral}, and \texttt{ArrayDeclaration}, both kinds of tests perform similarly, and snapshot tests have slightly higher medians (+0.8 to +5.7 points). For operator mutators, non-snapshot tests often have higher kill rates, especially for \texttt{UpdateOperator} (50.0\% vs.\ 17.0\%) and \texttt{ArithmeticOperator} (31.6\% vs.\ 1.5\%). However, these gaps are largely driven by small samples: the median repository has only two \texttt{UpdateOperator} mutants and five \texttt{ArithmeticOperator} mutants. The \texttt{ArithmeticOperator} difference is significant only before correction ($p=0.017$), and the \texttt{UpdateOperator} difference is not significant even before correction ($p=0.200$).

Overall, the kill rates by mutator type do not show that either kind of test specializes in certain mutations. The operator trend favors non-snapshot tests, but it is not reliable after correction. What distinguishes the two is which mutants each kind kills. As Figure~\ref{fig:mutation_score} shows, each kind kills mutants that the other misses.

\summarybox{\textbf{Answer to RQ2}}{
\textit{Combining snapshot tests with other tests detected more than \mutationEffect~times the number of mutants compared to tests without snapshot tests. While the mutant detection capability of snapshot tests is not significantly different from that of non-snapshot tests, the mutants they detect are different. This highlights the complementary nature of both testing methods. }
}

\section{Survey with Developers}\label{sec:qualitative}

This section details the design and results of a qualitative study conducted to address RQ3. We surveyed professional developers to understand the practical effectiveness, benefits, and challenges of snapshot testing.

\subsection{Survey Design}
\subsubsection{Overview}
We conducted a survey of developers to evaluate the effectiveness of snapshot testing through three dimensions: (1) the types of bugs detected, (2) the perceived benefits, and (3) the challenges encountered during the utilization. At the beginning of the survey, we asked developers about their general programming experience, their experience with snapshot testing, and the frameworks they use for snapshot testing. Following this, we presented a multiple-choice question to developers with experience in snapshot testing, asking them to identify the types of bugs they believe snapshot testing can prevent. To help categorize these bugs, we referred to the classification by Catolino\etal\cite{DBLP:journals/jss/CatolinoPZF19}. The details of these categories are described in Section~\ref{sec:survey-contents}. 
Additionally, we included open-ended questions to gather insights about the benefits of snapshot testing, notable experiences regarding bugs detected through snapshot testing, and the challenges developers have faced while utilizing snapshot testing.
Finally, we included an open-ended question inviting participants to share any tips they might have for using snapshot testing effectively.

\subsubsection{Questions}\label{sec:survey-contents}
Our survey form, created using Google Forms, consists of five sections: Demographic, Effectiveness, Benefits, Challenges, and Tips. Each section is described as follows:

\textbf{Demographic. }This section gathers information about the respondents, such as programming experience, confidence in their experience with snapshot testing, and the snapshot testing framework they use. Specifically, we asked the following three closed questions: 
\begin{enumerate}
    \item \textit{What is your programming experience?}\\
    Options: ``< 3 years'', ``3 - 5 years'', ``6-10 years'' and ``> 10 years''
    \item \textit{I'm experienced in snapshot testing.}\\
    Options: ``Strongly Agree'', ``Agree'', ``Neither Disagree Nor Agree'', ``Disagree'', and ``Strongly Disagree'' 
    \item \textit{What framework do you use for snapshot testing?}\\
    Options: ``Jest'', ``Chai with Mocha'', ``Vitest'', and ``Other (free texts)''
\end{enumerate}


\textbf{Effectiveness. } This section assesses the respondents' views on the effectiveness of snapshot testing in their development processes. Specifically, we asked about the types of issues that can be found by snapshot testing, using the following question:  
\begin{quote}
    \textit{Which types of issues do you think snapshot tests are good at detecting? }
\end{quote}
To facilitate easier responses, we provided predefined options based on the categories established by Catolino\etal\cite{DBLP:journals/jss/CatolinoPZF19} categorized bugs into several distinct types based on their root causes. These categories include Algorithmic Bugs, Concurrency Bugs, Configuration Bugs, Data Handling Bugs, Error Handling Bugs, GUI Bugs, I/O Bugs, Performance Bugs, and Security Bugs. Table~\ref{tab:bug_types} summarizes the details of these categories. We asked developers to select from these options (multiple selections allowed) and/or choose ``other'' to describe additional types of bugs in free text.

Note that we did NOT use a 5-level Likert
scale answer~\cite{chisnall1993questionnaire} (\ie ``Always'',``Often'', ``Sometimes'', ``Rarely'' and ``Never'') to ask whether snapshot testing is helpful for each category of bugs. This decision was made because asking developers to respond to each of the nine categories with 5-scale answers would impose a heavy burden on them. This could potentially decrease the number of responses, thereby hindering the collection of important insights from experts.

\begin{table}[t]
\small
\centering
\caption{Bug Categorization Made by Catolino\etal\cite{DBLP:journals/jss/CatolinoPZF19}}
\begin{tabular}{lp{9cm}}
\toprule
Bug Type & Description \\ \midrule
Algorithmic Bugs & Algorithmic bugs occur due to errors in the logic or algorithms used in the software. \\ \addlinespace
Concurrency Bugs & Concurrency bugs are related to issues that arise in multi-threaded or parallel execution environments. \\ \addlinespace
Configuration Bugs & Configuration bugs are caused by incorrect or suboptimal configuration settings. \\ \addlinespace
Data Handling Bugs & Data handling bugs involve errors in data processing, such as incorrect data formats or data corruption. \\ \addlinespace
Error Handling Bugs & Error handling bugs occur when the software fails to properly handle errors or exceptions. \\ \addlinespace
GUI Bugs & GUI bugs are related to issues in the graphical user interface, such as layout problems or unresponsive elements. \\ \addlinespace
I/O Bugs & I/O bugs are associated with input/output operations, including file handling and network communication. \\ \addlinespace
Performance Bugs & Performance bugs involve issues that degrade the performance of the software, such as memory leaks or inefficient algorithms. \\ \addlinespace
Security Bugs & Security bugs are vulnerabilities that can be exploited to compromise the security of the software. \\ \bottomrule
\end{tabular}
\label{tab:bug_types}
\end{table}

\textbf{Benefits. } This section explores the perceived benefits of using snapshot testing, including improvements in code quality, ease of maintenance, and other advantages. Specifically, we asked the following two open questions: 
\begin{enumerate}
    \item \textit{What are the major benefits of snapshot testing that motivate you to use it?}
    \item \textit{If you have some interesting experience in detecting certain bugs with snapshot tests, can you tell us more about it?}
\end{enumerate}

\textbf{Challenges. } This section identifies the challenges and obstacles respondents face when implementing snapshot testing, such as technical difficulties, learning curves, and integration issues. Specifically, we asked the following two open questions: 

\begin{enumerate}
    \item \textit{What problems or major difficulties do you face when using snapshot testing?}
    \item \textit{How would you like snapshot testing techniques to be improved? (If you haven’t encountered any difficulties or problems, please write ``none'')};
    \item \textit{``Do you have any tips on how to better use snapshot testing?}
\end{enumerate}


\subsubsection{Analysis}
To identify the benefits and challenges of snapshot testing, we manually examined the participants' responses. Specifically, we employed a card sorting approach~\cite{cardsorting}. Two authors first manually and independently examined each response, annotating the benefits and challenges. We categorized \bresponse responses for benefits and \cresponse for challenges. We then discussed each case to verify whether their labels expressed the same meaning, as free-text was used during the annotation. Note that the answers could be assigned multiple labels. 
Any conflict was resolved by discussing with a third author until agreement was reached. Subsequently, the three authors grouped similar labels, resulting in the final categories. The three annotators have 4, 13, and 15 years of programming experience. 

To facilitate the categorization process, we referred to the categories developed by a previous study\cite{DBLP:journals/jss/CruzRV23} that performed a grey literature review. If we found benefits and challenges that did not fit into their categories, we developed new categories. After the categorization process, the two authors reached an agreement in \benefitsAgreementRate\% of the inspected benefits and \challengesAgreementRate\% of the challenges, resulting in Cohen's kappa coefficients of \benefitsKappa~and \challengesKappa, respectively. These Cohen's kappa coefficients indicate \benefitsKappaInterpretation~and \challengesKappaInterpretation~levels of agreement, respectively~\cite{Landis1977TheMO}.

\subsection{Survey Participants}
In this survey, we invited developers with experience in snapshot testing to participate. To identify these developers, we first compiled a list of those who had made at least one change to the codebases of the repositories studied in Section~\ref{sec:study_design}. We then identified developers who had modified snapshot test methods at least once. This approach allowed us to survey a diverse group of developers who are experienced in snapshot testing.
Survey invitations were sent to \sent developers and were available for 30 days. Of these, \delivered emails were successfully delivered, while \undelivered emails were returned due to unreachable email addresses. In the end, we received \responses responses.

Table~\ref{tab:demographic} summarizes the demographics of the survey participants. 
Fifty participants (75.8\%) have more than 10 years of programming experience, ten participants (15.2\%) have 6-10 years, five participants (7.6\%) have 3-5 years, and one participant has less than 3 years. Additionally, 54 participants (81.8\%) expressed confidence in their experience with snapshot testing (\ie they answered ``Strongly agree'' or ``Agree'' to the statement ``I'm experienced in snapshot testing''). Six participants indicated insufficient knowledge of snapshot tests (\ie developers who chose ``Disagree'' or ``Strongly Disagree'' to the aforementioned statement). Therefore, \invalidresponses responses were excluded. Ultimately, we obtained \validresponses valid responses, resulting in a response rate of 5.2\%, which is comparable to the typical response rate of 5\% in most software engineering surveys~\cite{DBLP:journals/ese/LethbridgeSS05}.

\begin{table}[t]
\small
\caption{Demographics of the survey participants (N=66)}
\begin{tabular}{p{5cm}lrr}
\toprule
\textbf{Question} & \textbf{Options} & \textbf{Count} & \textbf{Percentage} \\
\midrule
\multirow{4}{5cm}{What is your programming experience?} 
& < 3 years & 1 & 1.5\% \\
& 3 - 5 years & 5& 7.6\% \\
& 6 - 10 years & 10 & 15.2\% \\
& > 10 years & 50 & 75.8\% \\
\midrule
\multirow{5}{5cm}{I’m experienced in snapshot testing.} 
& Strongly Agree & 23 & 34.8\% \\
& Agree & 31 & 47.0\% \\
& Neither Disagree Nor Agree & 6 & 9.1\% \\
& Disagree & 3 & 4.5\% \\
& Strongly Disagree & 3 & 4.5\% \\
\midrule
\multirow{4}{5cm}{What framework do you use for snapshot testing?} 
& Jest & 61 & 92.4\% \\
& Vitest & 20 & 30.3\% \\
& Chai with Mocha & 3 & 4.5\% \\
& Other (free texts) & 9 & 13.6\% \\
\bottomrule
\label{tab:demographic}
\end{tabular}
\end{table}

\subsection{Results}
In the following sections, we present the results summarizing the responses regarding the issues identified by snapshot testing, as well as its advantages and challenges. 
Note that the \faMale\ symbol represents direct quotes from survey participants.

\begin{table}[t]
\small
\caption{Issue types for which developers found snapshot testing effective (multiple selections allowed; \surveyRespondents{} respondents, \surveyTotalSelections{} selections in total)}
\begin{tabular}{lrrr}
\toprule
Issue type & \# & \% of respondents & \% of selections \\ \midrule
Program anomaly issue        & \rqFourProgramAnomalyCount & \rqFourProgramAnomalyPct\% & 31.3\% \\
GUI-related issue            & \rqFourGUICount            & \rqFourGUIPct\%            & 23.9\% \\
Test code-related issue      & \rqFourTestCodeCount       & \rqFourTestCodePct\%       & 16.4\% \\
Configuration issue          & \rqFourConfigCount         & \rqFourConfigPct\%         & 11.9\% \\
Performance issue            & \rqFourPerformanceCount    & \rqFourPerformancePct\%    & 2.2\% \\
Permission/deprecation issue & \rqFourPermissionCount     & \rqFourPermissionPct\%     & 2.2\% \\
Database-related issue       & \rqFourDatabaseCount       & \rqFourDatabasePct\%       & 2.2\% \\
Security issue               & \rqFourSecurityCount       & \rqFourSecurityPct\%       & 1.5\% \\
Network issue                & \rqFourNetworkCount        & \rqFourNetworkPct\%        & 0.7\% \\ \midrule
Other (free text)$^{*}$      & \rqFourOtherCount          & \rqFourOtherPct\%          & 7.5\% \\ \bottomrule
\multicolumn{4}{p{11cm}}{\footnotesize $^{*}$Strings, terminal UI output, matching nearly static responses, stylistic changes, detecting unintended public API surface changes, catching regressions, component rendering snapshot images, output of render methods and complex data structure tests, parsing big objects, etc.}
\end{tabular}
\label{tab:types}
\end{table}

\subsubsection{Issue Types Identified} We obtained \surveyRespondents{} responses concerning the types of issues identifiable via snapshot testing.
Table~\ref{tab:types} summarizes the number of issue types for which participants indicated that snapshot tests are useful. Since the responses could include multiple options, there are \surveyTotalSelections{} choices in total. Unless otherwise noted, percentages are relative to the \surveyRespondents{} respondents and therefore do not sum to 100\%. The most common issue was ``Program Anomaly issues,'' which include incorrect outputs, unexpected exceptions, and crashes due to logic issues. \rqFourProgramAnomalyCount{} participants (\ie \effectiveProgramAnomaly\%) indicated that snapshot testing is useful for this issue. The next most common issue was ``GUI-related issues,'' such as stylistic errors of GUI elements, which were reported by \rqFourGUICount{} participants (\effectiveGUIIssues\%). The popularity of these two issues is not surprising, as many grey literature sources have highlighted these benefits~\cite{DBLP:journals/jss/CruzRV23}.

However, surprisingly, the third most common issue was ``Test code-related issues,'' which include problems in test code due to (i) running, fixing, or updating test cases, (ii) intermittent tests, and (iii) the inability of a test to find de-localized bugs. We observed that \rqFourTestCodeCount{} participants (\effectiveTestCodeIssues\%) agreed with the effectiveness of this issue. 
Specifically, several developers indicate that snapshot tests can detect issues that other tests should have detected: 
\statement{Easy to spot unintended changes in snapshot data that may be missed with only basic unit test assertions}
\statement{It gives you a full picture of the output of something, rather than just testing for specific values. You might expect output x to change when input y changes, and therefore you test x. However, if you use snapshots you might catch that in fact z also changed when y changed. This would not have been caught in a normal test. Looking at the full output is also easier for a human to understand, since they might be able to judge whether it all makes sense as a whole.}

This statement implies that even though unit tests should detect these changes as errors, they did not work (\ie false negative). In other words, snapshot tests can find this kind of issue in unit tests. In particular, snapshot tests serialize many objects created during test exercises, which enables developers to identify what should have been tested, by knowing what values to be returned before and after a change.  

Additionally, the fourth most common category (\rqFourConfigCount{} participants, \rqFourConfigPct\%) was ``Configuration issues," which include building configuration files, local configuration files, external libraries, etc. For example, one developer mentioned that snapshot testing is useful for identifying changes caused by third-party library updates.

\statement{Catching issues on dependency upgrades and catching unexpected output}

\subsubsection{Advantages}
We categorized \bresponse responses regarding the benefits of snapshot testing into eight types. Five align with previous literature~\cite{DBLP:journals/jss/CruzRV23}, while three emerged as new findings in this study. The most common and second most common benefits of snapshot testing were ``Easy to implement'' and ``Prevent regression,'' which are consistent with the findings of the previous study. However, the third most common benefit was ``Easy to check results,'' where developers often appreciate that snapshot tests highlight the differences from previous behavior. The details of each category are described as follows.

\newcommand{\GraphAdvantage}{
    \begin{table}[t]
    \centering
        \caption{Survey results of advantage}
        \begin{tabular}{p{5.5cm}rr}\toprule
                            & \# &\%   \\ \midrule
        Easy to implement	& \implement & \implementPercentage \\
        Prevents regression	& \regression & \regressionPercentage \\
        \textbf{Easy to check results}	& \checkresult & \checkresultPercentage \\
        Serializable	& \seriarizable & \benefitSerializablePct\% \\
        \textbf{Easy to update}	& \update & \benefitEasyToUpdatePct\% \\
        Alternative to UI tests	& \AltUItest &  \benefitAlternativeToUITestsPct\% \\
        \textbf{Improve maintainability} & \maitainability & \benefitImproveMaintainabilityPct\% \\
        Easier and safer refactorings	& \refactoring & \benefitEasierRefactoringsPct\% \\\bottomrule
        \multicolumn{3}{l}{\footnotesize *Categories highlighted in bold were introduced in this study.}

        \end{tabular}
    \end{table}
}

\newcommand{\GraphDisadvantage}{
    \begin{table}[t]
    \centering
        \caption{Survey results of challenges}
        \begin{tabular}{p{9cm}rr}\toprule
                            & \# &\%   \\ \midrule
        Hard to review/interpret results	& \hardToReview & \challengeHardToReviewPct\% \\
        Introducing false positives	& \falsePositives &  \challengeFalsePositivesPct\% \\
        Unclear testing intention	& \unclearIntention & \challengeUnclearIntentPct\% \\
        \textbf{Easy to ignore issues and accidentally update snapshots}	& \easyToIgnore & \challengeEasyToIgnorePct\% \\
        Flaky behavior	& \flaky & \challengeFlakyPct\% \\
        \textbf{Lack of cross-platform capabilities}	& \crossPlatform & \challengeCrossPlatformPct\% \\
        \textbf{Increase maintenance cost}	& \maintainanceCost & \challengeMaintenanceCostPct\% \\
        Large/frequent snapshots reduce the interests of reviewers	& \largeSnapshots & \challengeLargeSnapshotsPct\% \\
        \bottomrule
        \multicolumn{3}{l}{\footnotesize *Categories highlighted in bold were introduced in this study.}

        \end{tabular}
        \label{tab:challenges}
    \end{table}
}

\GraphAdvantage 

\smallskip
\textbf{Easy to implement.}
This category originally comes from the grey literature review\cite{DBLP:journals/jss/CruzRV23}. We identified \implement~cases that fall into this category. 
Snapshot testing allows you to implement tests without writing extensive code across multiple lines, especially using only a single assertion method. Many developers mentioned that snapshot testing saves time in developing test code: 

\statement{This will help to reduce the time-consuming effort of writing assertion}
\statement{It saves time generating comparison data for tests that are either straightforward, and so not worth spending the time to write the expected outputs for the test, as one can just sanity check the snapshots, or for tests that require quite complex outputs, which take a lot of time to generate.}
\statement{When I need to write a large number of tests, I can use snapshots instead of manually describing objects, like using expect(testResult).toMatchInlineSnapshot() instead of expect(testResult). toEqual(veryLargeInlineObject)}

\smallskip
\textbf{Prevents regression.}
This category originally also comes from the grey literature review\cite{DBLP:journals/jss/CruzRV23}. We observed \regression~cases that fall into this category. 
Snapshot testing can detect subtle changes in snapshots before and after code modifications.
In particular, it allows detecting regressions in other teams' code caused by changes performed by developers from one team during multi-team development.\par

\statement{Snapshot testing is good for larger teams where one team changing a component may affect another team's code in unexpected ways.}

\smallskip
\textbf{Easy to check results. }Snapshot testing provides a view of results to show what different results are obtained after a change. 
This advantage of snapshot testing is helpful for developers when \textit{executing} snapshot tests. This category is newly found in this study and there are \checkresult~ instances that fall into this category. 
When snapshot tests are executed, the framework of snapshot testing compares two snapshots. If there are any differences between the two, the framework outputs the locations of the changes along with the content before and after the modifications. This view facilitates developers to understand what change happens in the production code. For example, developers can check the different behaviors on multiple browsers.
\statement{Snapshot testing (...) helps me to ensure that my library works consistently in different browsers, if snapshot test return same DOM snapshot (...)}

\smallskip
Additionally, many developers mentioned this advantage is useful for the review process because other team members can easily understand the change. 
\statement{test failures and behavior changes are straightforward in PR review.
}
\statement{3rd-party verifiable tests that can be shared with another code reviewer.}
\par

\smallskip
\textbf{Serializable. } Snapshot testing enables developers to check any serializable value. 
Different from ``Easy to check results'',
this advantage of snapshot testing is helpful for developers when \textit{writing} snapshot tests. 
This category originally comes from the grey literature review\cite{DBLP:journals/jss/CruzRV23} and \seriarizable~cases are observed.  
Snapshot tests can target anything that can be serialized. This allows for testing numbers, arrays, UIs, and large objects, giving them a broad range of applications. In particular, while writing unit tests manually for UIs or large objects can be very costly, snapshot tests can handle such cases with minimal additional cost.

\smallskip
\statement{It's an easy way to compare large chunks of data that would be costly to define in code. E.g. Output of API, etc.}
\statement{They can catch issues in UI elements that would otherwise not be explicitly tested for}

\smallskip
\textbf{Easy to update. }Once snapshot tests detect a change and developers admit the change is reasonable, snapshot test frameworks can update the asserted values automatically. For unit tests,  it is necessary to manually update the expected values in assertions when the behavior of the production code is modified. This category is newly introduced in this study and there are \update~ instances that fall into this category. 
This easy update function is appreciated by many developers: 

\statement{Automatically update assertions. Don't need to manually type out assertions and just verify them by looking at them.}\par
\statement{Quicker way to update and create test cases is the primary one.}

\smallskip
\textbf{Alternative to UI tests. }This category originally comes from the grey literature review\cite{DBLP:journals/jss/CruzRV23}. We found \AltUItest~ instances that fall into this category. 
Since snapshot tests can also be applied to UIs, they can be used in cases where you simply want to verify whether the UI is rendered correctly. In such scenarios, instead of running more costly UI tests, snapshot tests provide a lower-cost alternative for execution. Moreover, there are cases where the likelihood of detecting bugs increases.

\statement{Often act like end-to-end tests but are more reliable}

\smallskip
\textbf{Improve maintainability. }Snapshot testing helps developers keep clean code. This category is newly found in this study and there are \maitainability~ instances that fall into this category. 
In snapshot testing, the expected values of the assertions can be stored in a separate file called a snapshot file, rather than in the test file itself. This allows developers to separate the expected values from the test file, even if they are large, thereby maintaining the readability of the code.

\statement{The generated snapshot helps also to keep the testing code clean. }\par
\statement{Snapshot tests remove the need of having hardcoded values in your tests, e.g when testing that a component in particular state has the correct styling applied. It makes tests less brittle and more maintainable over time.}

\smallskip
\textbf{Easier and safer refactorings. }
This category originally comes from the grey literature review\cite{DBLP:journals/jss/CruzRV23}. We identified \refactoring~ instances that fall into this category. 
When a snapshot test passes, it indicates that the tested value has not changed, making refactoring easier. This is particularly useful in cases where refactoring involves only structural changes to the code, and it is required to ensure that the tested values remain unchanged.

\statement{Snapshot tests are also a quick way to ensure that the output remains the same when refactoring code.}
\statement{Being sure no unexpected changes happen during refactoring, bug fixing and so on.}


\subsubsection{Challenges}
We categorized \cresponse responses that contain the challenges of snapshot testing. Among them, \validRespondetDisadvantage~were valid responses. We identified eight major types of challenges: three types are newly introduced in this study, whereas five were identified by a previous grey literature review study\cite{DBLP:journals/jss/CruzRV23}. Table~\ref{tab:challenges} summarizes the challenges found in this study.

The most prevalent challenges of snapshot testing were ``Hard to review/interpret results'' and ``Introducing false positives.'' Although these challenges were not explicitly listed in the previous study, they may fall under the broader categories of ``Manual Verification,'' ``Large snapshots,'' and ``Fragility.'' We redefined these categories because the previous ones were too vague, making it unclear what specific issues developers were facing.

\GraphDisadvantage

\smallskip
\textbf{Hard to review/interpret results. }
A disadvantage of snapshot testing is that results can be difficult to interpret. Snapshots may become large, or the number of changing snapshots may grow, placing a significant burden on those who run tests and review code.

\statement{Snapshots can be too large, which makes them difficult to review, which leads to them being ignored or updated without thought}
\statement{Not always easy to review many snapshots for large PRs}

\smallskip
\textbf{Introducing false positives. }
This disadvantage of snapshot testing represents the frequent test failures even when no actual bugs exist. Snapshot tests fail if any part of the snapshot differs from the previous one, causing tests to fail even for intentional changes.
\statement{when we change some classname, it will break all related snapshots, even when they test some other part of component}
\statement{Sometimes whitespace is important, however when its not it can make the tests very noisy. Also when ordering is not important snapshot tests can be hard to work with.}

\smallskip
\textbf{Unclear testing intention. }
This disadvantage of snapshot testing represents the ambiguity of the test target. In snapshot testing, the scope of the test tends to become large, and when minor changes cause large or numerous snapshots to fail, it becomes difficult to identify the root cause.

\statement{Snapshots may be very difficult to understand, as they tend to be large and not informative.}
\statement{Snapshot tests, are too far away from the actual tests, and oftentimes capture too much, so they don't actually capture any intent.}

\smallskip
\textbf{Easy to ignore issues and accidentally update snapshots. }
This disadvantage of snapshot testing represents the risk of overlooking unintended changes and updating snapshots without noticing them. When a snapshot test fails due to changes in the codebase, the test executor can make the test pass by updating the snapshot. Since the update operation is automatically performed by the test framework, the update itself is simple. However, it is necessary to verify whether the update is truly appropriate. If the decision to update is not made carefully, it may introduce issues or render the test ineffective.
\statement{It's easy to miss a snapshot update that is actually relevant, especially if there are many updated snapshots.}
\statement{Easy to accidentally commit a change that wasn’t on purpose}

\smallskip
\textbf{Flaky behavior. }This category was also identified in the previous study. We observed this issue in \flaky~of the responses. Many developers complained that snapshot tests are sensitive to dynamically generated or varied values, such as IDs and timestamps, resulting in flaky behavior. 
\statement{(...) Another thing that often is a bit messy is ID's, e.g. snapshotting output of a database with an autogenerated ID, timestamp or similar, that changes on each snapshot.}
\statement{uuids, datetimes, anything non-deterministic that could change every snapshot}
\statement{Test results often depend on current time or current CPU performance or any other side effects. This challenge forces us to organize and prepare the testing dataset and preprocess the test results before matching them to the snapshot. I don't think we can improve it in any way.}

\smallskip
\textbf{Lack of cross-platform capabilities. }This category was not identified in the previous study. \crossPlatform~ respondents indicated that snapshot testing requires cross-platform support. Specifically, they mentioned that tests yield different results on various browsers, operating systems, and devices. One respondent advised that the new 'browser mode' feature in Vitest could help address this issue.

\statement{The most annoying part of snapshot testing is when it comes to test certain things that are different cross browsers and cross platforms.}
\statement{for image snapshots, the difference in rendering from the OS will impact the result, also sometimes we need to maintain masks.}

\smallskip
\textbf{Increase maintenance cost. } This category was newly identified in our study. We observed \maintainanceCost~cases where developers noted that maintaining snapshot test code incurs additional costs. 

\statement{I don't like the snapshots to be stored independent of the test as it makes the test less discoverable/understandable.}
\statement{(...) Snapshots always need to be maintained really well and usually a lot of additional formatters need to be built for testing.}

\smallskip
\textbf{Large/frequent snapshots reduce interests of reviewers. }
Two respondents warn that large or frequent snapshot updates reduce interests of reviewers. 

\statement{Snapshots can be too large, which makes them difficult to review, which leads to them being ignored or updated without thought}
\statement{(...) If snapshots are updated too often, perhaps it’s not the best option to use for that specific functionality, because reviewers are more likely to skim over these files, especially if they encounter them every PR.}

\smallskip
\textbf{Other. }We identified five cases that do not fit into the aforementioned categories. Each case has been assigned one of the following labels.

\begin{itemize}
    \item \textit{``Lower performance of IDEs due to large snapshot files''}
    \item \textit{``Introducing false negatives''}
    \item \textit{``Large effort for manual verification''}
    \item \textit{``False sense of security''}
    \item \textit{``Not sure if the snapshot is really true''}
\end{itemize}

\summarybox{\textbf{Answer to RQ3}}{
\textit{Developers value snapshot testing for its ease of implementation (\implementPercentage), regression prevention (\regressionPercentage), and straightforward result verification (\checkresultPercentage). It is perceived as effective at detecting Program Anomaly (\effectiveProgramAnomaly\%) and GUI issues (\effectiveGUIIssues\%). However, challenges such as difficulty reviewing large results (\hardToReviewRate\%) and frequent false positives (\falsePositiveRate\%) remain significant hurdles.}
}


\section{Implications}\label{sec:discussion}
This section discusses the broader implications of our findings for practitioners, tool developers, and researchers.

\subsection{Implications for Practitioners}
\textbf{Practitioners should adopt a hybrid testing strategy that combines snapshot and unit tests to maximize test effectiveness rather than relying on either method alone.}
Our quantitative analysis provides strong evidence that snapshot testing enhances test effectiveness when used alongside traditional unit tests. As shown in RQ1, the combination of snapshot and non-snapshot tests achieved significantly higher coverage than either method alone, with each approach covering distinct areas of production code. Similarly, RQ2 demonstrates that snapshot tests detect different types of mutants compared to unit tests, resulting in a higher overall mutation score when both testing styles are employed together. While a previous study~\cite{DBLP:conf/icsm/FujitaK0I23} observed that some projects use only snapshot tests, our findings suggest that this practice may create gaps in both test coverage and defect detection. We recommend practitioners adopt a complementary testing strategy: use unit tests for targeted assertions on critical logic and business rules, while leveraging snapshot tests to capture broader behavioral contracts and detect unexpected changes across the system.

\textbf{Practitioners should leverage snapshot testing beyond GUI testing to detect program anomalies, configuration issues, and test code problems.}
Snapshot testing is often perceived as a technique primarily for GUI testing. However, our survey results (RQ3) reveal that practitioners successfully apply it to diverse quality assurance scenarios. Respondents reported using snapshot tests most frequently for program anomalies (\effectiveProgramAnomaly\%), followed by GUI-related issues (\effectiveGUIIssues\%), test code issues (\effectiveTestCodeIssues\%), and configuration issues (\effectiveConfigureIssues\%). Notably, several developers indicated that snapshot tests can identify problems that basic unit test assertions might miss. As one participant explained, snapshot testing ``gives you a full picture of the output of something, rather than just testing for specific values,'' enabling developers to catch unexpected changes that would not be detected by traditional assertions. These diverse detection capabilities suggest that practitioners should consider applying snapshot testing to a wider variety of testing scenarios beyond UI verification.

\subsection{Implications for Tool Developers}
\textbf{Tool developers should prioritize addressing the difficulty of reviewing snapshot differences and cross-platform compatibility issues.}
Our survey results (RQ3) identify ``Hard to review/interpret results'' as the most prevalent challenge faced by practitioners (\challengeReviewingLargeResults\%). This finding aligns with the grey literature review by Cruz\etal\cite{DBLP:journals/jss/CruzRV23}, which also highlighted concerns about large and difficult-to-review snapshots. When snapshots become large or numerous, developers struggle to identify meaningful changes, often leading to snapshots being ``ignored or updated without thought,'' as one participant noted.

Additionally, we discovered that cross-platform compatibility is a concern that has not been previously documented in the literature. Four respondents specifically mentioned that snapshot tests produce different results across browsers, operating systems, and devices. One developer suggested that Vitest's ``browser mode'' feature could help address this issue. Tool developers should consider implementing features such as semantic diff highlighting, collapsible snapshot sections, and platform-aware snapshot comparison to reduce the cognitive burden on reviewers and improve cross-platform consistency.

\subsection{Implications for Researchers}
\textbf{Researchers should investigate how existing test quality improvement techniques can be adapted for snapshot testing to address flaky tests and false alarms.}
Our survey (RQ3) reveals that snapshot testing shares several challenges with traditional testing approaches. Specifically, \challengeFlakyPct\% of respondents encountered flaky behaviors, often caused by non-deterministic values such as timestamps and auto-generated IDs. As one developer noted, ``Test results often depend on current time or current CPU performance or any other side effects.'' Additionally, \challengeFalsePositivesPct\% of respondents identified false positives as a major challenge, where tests failed even for intentional changes. These issues have been extensively studied in the context of traditional test suites, and existing mitigation techniques~\cite{DBLP:journals/tosem/ParryKHM22}, such as flaky test detection and repair approaches, may be adaptable for snapshot testing contexts. Future studies should explore how these established solutions can be tailored to the unique characteristics of snapshot tests.

\textbf{Researchers may develop context-aware snapshot management techniques to help developers understand testing intentions and identify meaningful changes.}
We observed that many developers expressed difficulty understanding the intention behind snapshot tests when reviewing large snapshot files. As one participant noted, ``Snapshots may be very difficult to understand, as they tend to be large and not informative.'' Another developer mentioned that ``Snapshot tests are too far away from the actual tests, and oftentimes capture too much, so they don't actually capture any intent.'' This challenge is compounded by the ease with which developers can accidentally update snapshots without careful review, with \challengeEasyToIgnorePct\% of respondents reporting issues with unintentionally accepting incorrect changes. Future research should propose context-aware snapshot update prediction techniques that analyze code changes alongside snapshot modifications to help developers determine whether an update is intentional or indicates a potential defect, thereby reducing the cognitive burden of reviewing large snapshot differences.






\section{Threats to Validity}\label{sec:thread2validity}

In this section, we discuss potential threats to the validity of our study and the measures we took to mitigate them.

\subsection{Threats to Internal Validity}
\textit{Threats to internal validity} concern the factors we did not consider that might impact the causal relationships in our results. 
The manual classification of survey responses involved subjective interpretation that could introduce bias. To mitigate this threat, two researchers independently classified all responses. Their classifications yielded Cohen's kappa coefficients of \benefitsKappa~for benefits and \challengesKappa~for challenges, indicating \benefitsKappaInterpretation~and \challengesKappaInterpretation~agreement, respectively~\cite{Landis1977TheMO}. The agreement for challenges is only moderate, which suggests that these categories are harder to distinguish; to limit the resulting risk, all disagreements were discussed and resolved with a third researcher. However, we acknowledge that misinterpretation of participant responses cannot be entirely eliminated, particularly for ambiguous or brief answers. Besides, we utilized Stryker’s mutation operators, which may not inject all the fault types specific to JavaScript environments in practice. Although alternative operators could influence the observed effectiveness of snapshot versus unit testing, Stryker represents the industry standard and addresses the most frequent fault patterns.

\subsection{Threats to Construct Validity}
\textit{Threats to construct validity} concern the relation between theory and observation. For the measurement of coverage and mutation scores, we used the most basic command ``\texttt{npx jest --coverage}'' and ``\texttt{stryker init}''. While this ensures consistency and reproducibility, it may not reflect the optimal testing configurations used by each project. 
Identifying and executing project-specific test configurations for hundreds of projects would require substantial manual effort and introduce additional variability. Our standardized approach provides a fair comparison across projects. Moreover, we used code coverage and mutation scores as proxies for defect detection capability. High coverage does not guarantee fault detection, and mutation testing may not represent real-world defects. However, these metrics are widely accepted in software testing research and provide quantifiable and objective measures for comparative analysis.

\subsection{Threats to External Validity}
\textit{Threats to external validity} concern the generalizability of our findings. Our study leverages \texttt{package.json} to identify repositories that employ Jest. This file is generated by popular package managers like NPM and YARN, and other tools might not produce this file. However, as NPM and YARN are the most popular package managers for JavaScript and TypeScript, we expect the number of missed repositories to be small and unlikely to significantly affect our conclusions. Another threat is that our findings are specific to the JavaScript/TypeScript ecosystem and the Jest testing framework. Snapshot testing implementations in other languages and frameworks may exhibit different characteristics. 

\section{Related Work}\label{sec:related_work}
This section introduces related work on snapshot testing and test quality.

\subsection{Snapshot Testing}
While various types of tests including unit tests, integration tests, and GUI tests have been extensively studied by researchers~\cite{DBLP:journals/tse/AnicheTZ22}\cite{DBLP:journals/ese/SpadiniABB19}\cite{DBLP:conf/kbse/TufanoPBPOLP16}\cite{DBLP:conf/icsm/ChristopheSRM14}, only a few of them~\cite{DBLP:conf/msr/BuiR23}\cite{DBLP:conf/icsm/FujitaK0I23}\cite{DBLP:journals/jss/CruzRV23} have focused on snapshot testing. 
Bui\etal\cite{DBLP:conf/msr/BuiR23} created a dataset of repositories on GitHub that use snapshot tests. They extracted projects using Jest, one of the frameworks that implement snapshot testing, based on GitHub tags, and selected repositories containing files with the snap extension to create the dataset. This utilizes the fact that when snapshot testing is executed by Jest, files with the snap extension are created. Their dataset includes a variety of artifacts such as the number of snap files and test files collected from 686 repositories that use snapshot testing. 

Fujita\etal\cite{DBLP:conf/icsm/FujitaK0I23} conducted an empirical investigation into the adoption and evolution of snapshot testing. Unlike the approach by Bui \etal\cite{DBLP:conf/msr/BuiR23}, which identified repositories using the ``jest'' tag, they first collected GitHub repositories with over 1,000 stars and then examined their package files and directories to confirm the use of Jest and snapshot files. Their study of 569 repositories using snapshot testing revealed that projects employing snapshot testing have more than 1.6 times the number of test cases compared to those that do not. Additionally, updates to snapshots often coincide with changes in JavaScript, TypeScript, and test files.

Cruz \etal\cite{DBLP:journals/jss/CruzRV23} investigated the practical adoption of snapshot testing through a grey literature review. They analyzed data from various industry sources to uncover the benefits and challenges associated with snapshot testing. The authors conducted a Google search, which returned 18,600 documents. From these, they selected and reviewed 50 documents, including blog posts, technical articles, and tool documentation, to gather insights from practitioners. Their findings indicate that while snapshot testing is praised for its simplicity and effectiveness in preventing regressions, it also poses challenges such as fragility and maintenance overhead. While their study shares a similar goal, our quantitative analysis and practitioner survey strengthen the findings.

\subsection{Test Effectiveness and Code Coverage}

Kochhar\etal\cite{DBLP:conf/wcre/KochharTL15} analyzed the correlation between the size of tests and coverage with test effectiveness using the history of actual large-scale system development. The analysis of two systems showed that while the correlation between test size and test effectiveness was weak for one and strong for the other, test coverage had a moderate correlation with test effectiveness for one system and a strong correlation for the other.

Kochhar\etal\cite{DBLP:journals/tr/KochharLLN17} conducted a large-scale empirical study to examine the correlation between code coverage and the number of real post-release defects in 100 open-source Java projects drawn from GitHub and tracking bugs via JIRA. Their findings showed that, at the project level, code coverage had an insignificant correlation with the number of post-release bugs, or the number of bugs normalized by Lines of Code (LOC) or cyclomatic complexity. However, the coverage normalized by complexity (Coverage/Complexity) had a moderate negative correlation with the total number of post-release bugs at the project level.

Chekam\etal\cite{DBLP:conf/icse/ChekamPTH17} investigated the relationship between coverage and defect detection capability, paying attention to the timing and type of coverage acquisition. 
The results showed that mutation coverage is most related to defect detection capability, and the failure detection rate increases significantly only when coverage reaches a high level.

Ivankovic\etal\cite{DBLP:conf/sigsoft/IvankovicPJF19} described Google’s continuous code coverage infrastructure, detailing how the computed code coverage information is visualized, used, and integrated into the developer workflow at scale. 
Their study demonstrated that while automated coverage computation is crucial (only 0.5\% of developers manually invoke coverage daily), successfully scaling such an infrastructure requires considerable multi-year effort to address technical challenges, particularly those related to dealing with failed coverage computations. 
Usability and low overhead are the key factors for achieving positive developer sentiment regarding code coverage.

Ivankovic\etal\cite{DBLP:conf/icse/IvankovicPKLKJ024} introduced Productive Coverage, which addresses the limitation that traditional coverage treats all test goals equally. By identifying uncovered code similar to well-tested code or frequently executed in production, the approach directs developers toward critical testing gaps. Evaluation across C++, Java, Go, and Python showed strongly positive developer sentiment and measurable improvements in test quality and code-review efficiency.

\subsection{Test Quality and Mutation Testing}
While mutation testing is computationally expensive, research at Google by Petrovic \etal \cite{DBLP:conf/icse/PetrovicI18} demonstrates that costs can be mitigated by filtering for ``useful'' mutants that correlate with real-world bugs. This correlation is further supported by Just \etal \cite{DBLP:conf/sigsoft/JustJIEHF14}, who found that mutation scores are a more reliable predictor of bug-detection capability than standard statement coverage. Similarly, Perretta\etal\cite{DBLP:conf/issta/PerrettaDG022} found a strong correlation between the evaluation by mutation testing and the evaluation by manually created tests when judging whether the code written by programming students was correct. To assess whether the test suites created by the students functioned well, teachers manually created tests and used mutation testing. 




Mirshokraie\etal\cite{DBLP:conf/icst/MirshokraieMP13} proposed MUTANDIS, a mutation testing technique for JavaScript that combines static and dynamic analysis to guide mutant generation toward non-equivalent mutants. Their approach ranks functions using a metric called FunctionRank, which is based on dynamic call frequency, and selects variables for mutation according to their impact on program behavior. Their evaluation demonstrated that 93\% of the generated mutants were non-equivalent. 

Rodríguez-Baquero and Linares-Vásquez \cite{DBLP:conf/issta/Rodriguez-Baquero18} proposed Mutode, a generic mutation testing tool for JavaScript and Node.js applications. They implemented 43 mutation operations grouped into 16 mutators, covering various aspects from boolean literals to function removals. The tool was evaluated on 12 of the top 20 most depended-upon npm modules, generating an average of 2,952 mutants per module and achieving an average mutation coverage of 70.59\%. The results demonstrated that Mutode can effectively identify surviving mutants that reveal gaps in test suites, providing developers with actionable insights for test improvement in JavaScript ecosystems.

Gopinath\etal\cite{DBLP:conf/icse/GopinathJG14} investigated which code coverage criteria best predict mutation detection capability from a developer's perspective. The study analyzed 318 real-world Java projects from GitHub, using both manually created test suites and automatically generated suites. The experiment compared various coverage metrics (statement, branch, block, and path coverage) against mutation scores as a quality indicator. The results showed that statement coverage provides the best prediction of mutation kills, 
outperforming branch coverage which previous research had suggested as superior.

\section{Conclusion}\label{sec:conclusion}

In this study, we presented a comprehensive empirical investigation into the effectiveness of snapshot testing in front-end development, utilizing both quantitative data from \studiedrepo GitHub repositories and qualitative insights from a survey of \validresponses experienced practitioners. 
Our mixed-methods analysis reveals that integrating snapshots significantly boosts median statement coverage to \coverageAllMedian\% and increases mutant detection by over \mutationEffect~times. Practitioners appreciate snapshot testing for its low implementation cost and its efficacy in catching Program Anomaly (\effectiveProgramAnomaly\%) and GUI (\effectiveGUIIssues\%) defects, which can often identify issues that standard assertions miss. However, snapshot testing also faces significant challenges, including high cognitive load during reviews, false positives during intentional updates, and flakiness caused by non-deterministic values (\eg timestamps).
Future work should focus on developing techniques to improve the reviewability of snapshot differences and mitigate the fragility associated with large snapshot files.

\begin{acks}
We gratefully acknowledge the financial support of JSPS for the KAKENHI grants (JP24K02921, JP24K02923), the Bilateral Program grant (JPJSBP120239929), as well as JST for the PRESTO grant (JPMJPR22P3), the ASPIRE grant (JPMJAP2415), and the AIP Accelerated Program (JPMJCR25U7).
\end{acks}

\balance
\bibliographystyle{ACM-Reference-Format}
\bibliography{800_Bib/references,800_Bib/mutants}


\end{document}